\documentclass[acmtoms]{acmtrans2m}
\newdef{definition}[theorem]{Definition}
\newdef{remark}[theorem]{Remark}

\usepackage{amsmath}
\usepackage{amssymb}
\usepackage{bm}
\usepackage{url}
\usepackage[only,llbracket,rrbracket]{stmaryrd}
\usepackage{graphicx}

\newcommand{\vect}[1]{\bm{#1}}
\newcommand{\dx}{\, \mathrm{d}x}
\newcommand{\dX}{\, \mathrm{d}X}
\newcommand{\ds}{\, \mathrm{d}s}

\newcommand{\unorm}[1]{%
  \left\vert\kern-0.9pt\left\vert\kern-0.9pt\left\vert #1
    \right\vert\kern-0.9pt\right\vert\kern-0.9pt\right\vert}

\newcommand{\brac}[1]{\left( {#1} \right)}
\newcommand{\bracc}[1]{\left\{ {#1} \right\}}

\newcommand{\jump}[1]{\llbracket {#1} \rrbracket}
\newcommand{\avg}[1]{\langle {#1} \rangle}

\newcommand{\ffc}{FFC}
\newcommand{\dolfin}{DOLFIN}
\newcommand{\fenics}{FEniCS}
\newcommand{\ufc}{UFC}

\usepackage{verbatim}
\newenvironment{code}[1]%
{\center\tabular{c}\hline\\ \footnotesize\minipage{#1\textwidth}\verbatim}
{\endverbatim\endminipage\\ \\ \hline\endtabular\endcenter}

\title{Optimisations for quadrature representations of finite element tensors
       through automated code generation}
\author{KRISTIAN B. {\O}LGAARD \\
        Faculty of Civil Engineering and Geosciences \\
        Delft University of Technology
\\ and \\
        GARTH N. WELLS \\
        Department of Engineering \\
        University of Cambridge}
\begin{abstract}
We examine aspects of the computation of finite element matrices and
vectors which are made possible by automated code generation. Given
a variational form in a syntax which resembles standard mathematical
notation, the low-level computer code for building finite element
tensors, typically matrices, vectors and scalars, can be generated
automatically via a form compiler.  In particular, the generation of
code for computing finite element matrices using a quadrature approach
is addressed.  For quadrature representations, a number of optimisation
strategies which are made possible by automated code generation are
presented.  The relative performance of two different automatically
generated representations of finite element matrices is examined, with
a particular emphasis on complicated variational forms.  It is shown
that approaches which perform best for simple forms are not tractable
for more complicated problems in terms of run time performance, the
time required to generate the code or the size of the generated code.
The approach and optimisations elaborated here are effective for a range
of variational forms.
\end{abstract}


{}
\begin{document}
\begin{bottomstuff}
  K.B.~{\O}lgaard,
  Faculty of Civil Engineering and Geosciences,
  Delft University of Technology
  Stevinweg 1, 2628~CN Delft, Netherlands.
  Email: \url{k.b.oelgaard@tudelft.nl}.
  \newline
  G.N.~Wells, Department of Engineering, University of Cambridge,
  Trumpington Street, Cambridge CB2 1PZ, United Kingdom.
  Email: \url{gnw20@cam.ac.uk}.
\end{bottomstuff}
%------------------------------------------------------------------------------
\maketitle
%------------------------------------------------------------------------------
\section{Introduction}
The rapid development of solvers for a variety of partial differential
equations while achieving optimal or near-optimal run time performance
is a possibility offered by automated computer code generation. Rapid
development and high performance can be reconciled by introducing
a compiler that translates high-level mathematical representations
of variational forms into low-level computer code. The \fenics{}
Form Compiler (henceforth \ffc{}) is an example of one such
compiler~\cite{ffc:www,kirby:2006c}.  \ffc{} takes as input a
variational form, posed in a high-level mathematical language,
and generates code for the computation of the element tensors
(element matrices, vectors or scalars) in a low-level language,
such as C++. The generated code serves as input to a finite element
assembler. \ffc{} version 0.5.1 generates C++ code consistent with
the \ufc{} specification~\cite{ufc:manual,alnaes:2008}, and therefore
the generated code can be used by any assembly library which supports
the specification.  DOLFIN~\cite{logg:2009,dolfin:www} is an example
of an assembly library which supports the \ufc{} specification.  The
developments presented in this work are implemented in FFC.  A related
effort is the SyFi Form Compiler (SFC) \cite{syfi:www,alnaes:2009}.
Like FFC, SFC generates low-level code for a high-level input, but
follows a different philosophy in relying on a symbolic engine.

The automated generation of computer code for finite element tensors
provides scope for various representations and optimisations which
are not feasible via conventional code development approaches. A
possibility is to adopt a `tensor contraction' representation
of element tensors, rather than the classical quadrature-loop
representation \cite{kirby:2006c,oelgaard:2008}.  The approach is
based on the multiplicative decomposition of an element tensor into
two tensors, one of which depends only on the differential equation
and the chosen finite element bases and can therefore be computed
prior to run time.  It has been proved for classes of problems
that the tensor contraction representation is more efficient than
the traditional quadrature approach, and the speed-ups can be quite
dramatic \cite{kirby:2006c}.  Furthermore, strategies which analyse the
structure of the tensor contraction representation can yield improved
performance \cite{kirby:2005,kirby:2006b}.  However, in contrast to
the quadrature-loop approach, the tensor contraction representation is
somewhat specialised as it cannot be extended trivially to non-affine
isoparametric mappings while maintaining efficiency, and it is
not effective for classes of nonlinear problems which require the
integration of functions that do not come from a finite element space
\cite{automated_solids:2008}. The attractive feature of the approach is
the run time performance for classes of problems.

It has been our experience that the tensor contraction approach does
not scale well for moderately complicated and complicated forms. This
is manifest in three ways: the time required to generate low-level code
for a variational form becomes prohibitive or may fail due to memory
limitations or limitations of underlying libraries; the size of the
generated code is such that the compilation of the generated low-level
code is prohibitively slow and file size limitations of compilers acting
on the low-level code may be exceeded; and the run time performance
deteriorates rapidly relative to a quadrature approach.  Complicated
forms are by no means exotic. Many common nonlinear equations, when
linearised, result in forms which involve numerous function products.
It was when addressing these types of problems that we found automated
code generation using the tensor contraction representation would
frequently break down.  Factors that determine the complexity of a form
are the number of coefficient functions, the number of derivatives and the
polynomial orders of the finite element basis functions.  Approaches to
reduce the time required for the code generation phase when using the
tensor contraction representation have been developed and implemented in
\ffc{} \cite{kirby:2006d}, although these cannot counter the inherently
expensive nature of the approach for complicated forms.  Various issues
with automated code generation, particularly scaling, are only borne out
when considering complicated forms.  Naturally, automated code generation
is most appealing when considering complicated variational forms which are
time consuming to program, difficult to optimise and problematic to debug.

We address here issues pertinent to automated code generation for
quadrature representations of finite element tensors, and in particular
optimisations which are made possible by automation and could not be
reasonably expected of a developer to program `by hand'.  We wish in
particular to target complicated forms, for which the tensor contraction
approach performs poorly.  In assessing the tensor contraction and
quadrature representations, we consider
\begin{enumerate}
  \item The run time performance of the generated code;
  \item The size of generated code; and
  \item The speed of code generation phase.
\end{enumerate}
The relative importance of these points may well shift during a
development cycle.  During initial development, it is likely that the
speed of the code generation phase and the size of the generated code
are most important, whereas at the end of the development cycle run time
performance is likely to be the most crucial consideration.  As we will
show, there is typically a correlation between the three points.

All developments which we present are implemented in \ffc{}, which is
freely available at \url{http://www.fenics.org/} under the GNU Public
License.  \ffc{} is a component of the FEniCS project~\cite{fenics:www},
which consists of a suite of tools which aim to automate computational
mathematical modelling, all of which are released under a GNU public
license. The examples presented in this work can be `compiled' using
\ffc{} version~0.5.1.

The remainder of this work is arranged as follows. We summarise
automated code generation and representations of finite element
tensors in Section~\ref{sec:automated_code_gen}. Then, we describe in
Section~\ref{sec:a_priori_optimisations} the optimisations for quadrature
representations which we have applied.  Examples and benchmarks on the
performance of quadrature and tensor contraction representations are
presented in Section~\ref{sec:benchmarks}, after which conclusions are
drawn in Section~\ref{sec:conclusions}.
%------------------------------------------------------------------------------
\section{Finite element tensors and automated computer code generation}
\label{sec:automated_code_gen}
\subsection{Overview}
We review briefly in this section two representations, quadrature
and tensor contraction, of an element stiffness matrix. We choose as a
canonical example the bilinear form corresponding to the weighted Laplace
equation $-\nabla \cdot (w \nabla u) = 0$, where $u$ is unknown and $w$
is prescribed. The bilinear form associated with the variational form
of the weighted Laplacian reads
\begin{equation}
  a\brac{v, u} = \int_{\Omega} w \nabla v \cdot \nabla u \dx.
\label{eq:weightedlaplacian}
\end{equation}
We assume that all functions in the form come from the finite element
space
\begin{equation}
  V = \bracc{v \in H^{1}\brac{\Omega}: \ v\vert_K \in P_{k}\brac{K}
   \forall K \in \mathcal{T}},
\label{eq:space_H1}
\end{equation}
where $P_{k}\brac{K}$ denotes the space of Lagrange polynomials of degree
$k$ on the element $K$ of the standard triangulation of $\Omega$, which
is denoted by~$\mathcal{T}$.  The local element matrix, often known as
the `stiffness matrix', for the cell $K$ is given by
\begin{equation}
  A^{K}_{i_{1} i_{2}} = \int_{K} w \nabla \phi^{K}_{i_1} \cdot \nabla \phi^{K}_{i_2} \dx,
\label{eq:localtensor}
\end{equation}
where $\bracc{\phi^{K}_{i}}$ are the local basis functions which span
$V_{h}$ on the element~$K$.

The task of the form compiler is to take an input which resembles
the notation of equation~\eqref{eq:localtensor} and generate
low-level code.  The \ffc{} input for this problem is shown in
Figure~\ref{fig:weightedlaplacian_input} for continuous piecewise cubic
functions on tetrahedra as a basis for all functions in the form.
\begin{figure}
\begin{code}{0.8}
element = FiniteElement("Lagrange", "tetrahedron", 3)

v = TestFunction(element)
u = TrialFunction(element)
w = Function(element)

a = w*dot(grad(v), grad(u))*dx
\end{code}
\caption{\ffc{} input for the weighted Laplacian form on cubic tetrahedral elements.}
\label{fig:weightedlaplacian_input}
\end{figure}
Computer code is generated from the input shown in
Figure~\ref{fig:weightedlaplacian_input} by simply running the compiler
\ffc{} on the input code. Options can be provided which affect aspects
of the generated code.  Relevant to the topic of this work are the
representation options `\texttt{-r quadrature}' for quadrature
representation and `\texttt{-r tensor}' for tensor contraction
representation, both of which are summarised in this section.  \ffc{}
generates not only the code for computing the element matrix $A^{K}_{i_{1}
i_{2}}$, but also a degree-of-freedom mapping for use in assembly,
as well as a number of utility functions, such as code for evaluating
finite element functions at arbitrary points.  The code generated by
FFC conforms to the \ufc{} specification, and can be used to assemble
global matrices and vectors by an assembler which supports the \ufc{}
specification, such as the \fenics{} component \dolfin{}, which is a
problem solving environment responsible for assembling the global system
and solving the arising linear system of equations.

\ffc{} implements support for basic differential and algebraic
operators.  The operators which are used in the examples in this work
are: the gradient, \texttt{grad(v)}; the divergence, \texttt{div(v)};
multiplication \texttt{mult(w, v)}; and inner products \texttt{dot(v, u)}.
Different types of integrals are also available. Presented examples
will use integration over a cell, \texttt{*dx}, and integration over
interior facets, \texttt{*dS}. The latter will be used in a discontinuous
Galerkin example. Related to discontinuous Galerkin methods, the compiler
offers the possibility of restricting functions evaluated on facets
to either the plus side or the minus side of a given facet, which is
expressed as \texttt{v('+')} and \texttt{v('-')}, respectively. Also
relevant to discontinuous Galerkin methods are operators for the jump,
\texttt{jump(v)}, and the average, \texttt{avg(v)}, of a function on a
cell facet.  Mixed elements with arbitrary combinations of functions
and function spaces are supported (see \citeN{logg:2009} for some
examples).  \ffc{} is built on top of Python, and therefore inherits
Python syntax. This makes the addition of user-defined operators simple,
and in combination with the language of \ffc{}, makes it possible to
define a wide range of variational forms simply and compactly.

\ffc{} itself does not generate basis functions, but relies on the library
FIAT \cite{kirby:2004} for the computation of basis functions and the
generation of quadrature schemes.  Exact quadrature is used by default.
\ffc{} computes the polynomial order of the form and uses a scheme
based on the Gauss-Legendre-Jacobi rule mapped onto simplices (see
\cite{karniadakis:book} for details of such schemes).  The requested
quadrature scheme is computed by FIAT.  This means that for exact
integration of a second-order polynomial, \ffc{} will use two quadrature
points in each spatial direction i.e., $2^3 = 8$ points per cell in
three dimensions.  \ffc{} does provide an option for a user to specify
the number of quadrature points, thereby permitting inexact quadrature.

%------------------------------------------------------------------------------
\subsection{Quadrature representation}
\label{subsec:quadrature}
Finite element codes typically deploy quadrature at run time for the
numerical integration of local element tensors.  Assuming the same local
basis for all functions in equation~\eqref{eq:localtensor}, as indicated
in the compiler input in Figure~\ref{fig:weightedlaplacian_input}, and
a standard affine mapping $F_K : K_0 \rightarrow K$ from a reference
element~$K_{0}$ to any given element $K \in \mathcal{T}$ (recalling that
$\mathcal{T}$ is the triangulation of the domain of interest $\Omega$),
a quadrature scheme for equation~\eqref{eq:localtensor} reads
\begin{equation}
\small
  A^K_{i_{1} i_{2}}
  =
  \sum_{q=1}^{N}
  \sum_{\alpha_{3}=1}^n
  \Phi_{\alpha_{3}}(X^q)
  w_{\alpha_{3}}
  \sum_{\alpha_1=1}^d \sum_{\alpha_2=1}^d
  \sum_{\beta=1}^d
  \tfrac{\partial X_{\alpha_1}}{\partial x_{\beta}}
  \tfrac{\partial \Phi_{i_1}(X^q)}{\partial X_{\alpha_1}}
  \tfrac{\partial X_{\alpha_2}}{\partial x_{\beta}}
  \tfrac{\partial \Phi_{i_2}(X^q)}{\partial X_{\alpha_2}}
  \det F_K'
  W^q,
\label{eq:weightedlaplacian,quadraturerepresentation}
\end{equation}
where a change of variables from the reference coordinates $X$ to the real
coordinates $x = F_K(X)$ has been used. In the above equation, $N$ denotes
the number of integration points, $d$ is the dimension of $\Omega$,
$n$ is the number of degrees of freedom for the local basis of~$w$,
$\Phi_{i}$ denotes basis functions (shape functions) on the reference
element, and $W^q$ is the quadrature weight at integration point~$X^q$.
Since it is assumed that the function $w$ comes from a finite element
space, or is interpolated in a finite element space, it is necessary to
loop over the degrees of freedom associated with~$w$. The form compiler
also supports `quadrature functions' for which functions are evaluated
directly at quadrature points and to which no function space is attached.
The use of such functions is essential for a wide range of nonlinear
problems~\cite{automated_solids:2008}.  However, if quadrature functions
are used, the computed element matrices and vectors may differ from
those computed using the tensor contraction approach which is presented
in the following section. Therefore we will assume in all examples that
coefficient functions are interpolated in a finite element space.

%------------------------------------------------------------------------------
\subsection{Tensor contraction representation}
\label{sec:tensor}
In reviewing the tensor contraction representation
approach, we follow the work of~\citeN{kirby:2006c}.  Taking
equation~\eqref{eq:weightedlaplacian,quadraturerepresentation} as the
point of departure, the tensor contraction representation of the element
matrix for the weighted Laplacian is expressed as
\begin{equation}
  A^K_{i_{1} i_{2}} = \sum_{\alpha_1=1}^d \sum_{\alpha_2=1}^d \sum_{\alpha_3=1}^n
  \det F_K' w_{\alpha_3}
  \sum_{\beta=1}^d
  \frac{\partial X_{\alpha_1}}{\partial x_{\beta}}
  \frac{\partial X_{\alpha_2}}{\partial x_{\beta}}
  \int_{K_0} \Phi_{\alpha_3}
  \frac{\partial \Phi_{i_1}}{\partial X_{\alpha_1}}
  \frac{\partial \Phi_{i_2}}{\partial X_{\alpha_2}}
  \dX.
\label{eq:weightedlaplacian,tensorrepresentation}
\end{equation}
Noteworthy is that the integral appearing in
equation~\eqref{eq:weightedlaplacian,tensorrepresentation} is independent
of the cell geometry and can therefore be evaluated prior to run time.
The remaining terms, with the exception of $w_{\alpha_{3}}$, depend
only on the geometry of the cell. Exploiting this observation, the
element tensor $A^{K}_{i_{1} i_{2}}$ can then be expressed as a tensor
contraction,
\begin{equation}
  A^K_{i_{1} i_{2}} = \sum_{\alpha} A^0_{{i_{1} i_{2}}\alpha} G_K^{\alpha},
\label{eq:tensorrepresentation}
\end{equation}
where the tensors $A^{0}_{{i_{1} i_{2}}\alpha}$ (the `reference tensor')
and $G_K^{\alpha}$ (the `geometry tensor') are defined as
\begin{align}
  A^{0}_{{i_{1} i_{2}}\alpha}
      &= \int_{K_0} \Phi_{\alpha_3} \frac{\partial \Phi_{i_1}}{\partial X_{\alpha_1}}
          \frac{\partial \Phi_{i_2}}{\partial X_{\alpha_2}} \dX,
\\
  G_K^{\alpha}
      & = \det F_K' w_{\alpha_3} \sum_{\beta=1}^d
          \frac{\partial X_{\alpha_1}}{\partial x_{\beta}}
          \frac{\partial X_{\alpha_2}}{\partial x_{\beta}}.
\end{align}
We refer to~\citeN{kirby:2006d} for a generalisation of the approach.

Using \ffc{} to generate computer code for the tensor
contraction representation, the reference tensor $A^{0}_{{i_{1}
i_{2}}\alpha}$ is precomputed and the contraction in
equation~\eqref{eq:tensorrepresentation} is unrolled.  By unrolling
the contraction any zero-valued entry of the reference tensor can be
detected during the code generation stage and the corresponding code can
therefore be omitted.  For a certain class of simple forms this can lead
to a tremendous speed-up when evaluating the element matrices relative
to a quadrature approach \cite{kirby:2006c}.  Note, however, that as
the number of functions and derivatives present in the variational form
increases, the rank of both the reference tensor and the geometry tensor
increases, thereby increasing the complexity of the tensor contraction.
%------------------------------------------------------------------------------
\section{A priori optimisations for quadrature representation}
\label{sec:a_priori_optimisations}

The automated generation of code provides scope for employing
optimisations which may not be practically feasible in hand-generated
code.  An example of such an approach which is pertinent to
the tensor contraction representation involves the analysis of
structures in the reference tensor in order to minimise the number
of floating point operations required to compute an element matrix or
vector~\cite{kirby:2005,kirby:2006b,kirby:2008}.  For simple problems,
this can lead to a significant reduction in the number of operations
required to compute an element tensor.  However, it is our experience
that one is generally not well-rewarded for sophisticated optimisation
strategies.  Such strategies may not scale well in terms of the required
computer time to perform the optimisations for moderately complex
variational forms and prove to be prohibitive in terms of time and memory.
This is in conflict with the goal of minimising development time (code
generation phase), as described in the Introduction.  Our experience
indicates that simple optimisations, some of which are described in this
section, offer the greatest rewards, even to the extent that the cost of
evaluating element tensors becomes negligible relative to other aspects
of a computation, such as insertion of entries into a sparse matrix.

We outline here three simple \emph{a priori} approaches for optimising
generated code for the quadrature representation of an element tensor.
The strategies are implemented in FFC.  The central idea of all three
methods is to implement low-cost strategies to reduce the number of
floating point operations required to evaluate the local element tensor.
By low-cost optimisations, we imply strategies which do not impact the
time required for the code generation phase adversely.  The optimisations
which we have implemented are:
\begin{longenum}
\item Tabulation of basis functions and basis function derivatives:
Basis functions are evaluated and tabulated at integration points in the
generated code (although the compiler also generates functions for the
evaluation of basis functions and their derivatives at arbitrary points
at runtime).
\item Eliminate floating point operations on zeros: Basis functions and
derivatives of basis functions that are zero-valued at all integration
points may be identified and eliminated during the generation phase,
thereby reducing the dimension of the loops concerning these functions.
This process is comparable to dead-code elimination in compiler
jargon.  In particular, when taking derivatives of basis functions on a
reference element, or in the case of mixed elements, zeros often appear.
This requires the creation of a map for indices in order to correctly
access the basis values. This mapping results in memory not being
accessed contiguously at run time and can therefore potentially lead to a
performance drop, but in our experience this effect is outweighed by the
reduction of operations.  The elimination of operations on zero terms is
similar to the strategy that the tensor contraction representation applies
when unrolling the code.  The major difference being that the quadrature
representation can only eliminate contributions that are zero for all
quadrature points, unlike the tensor contraction representation which
can eliminate all zero-valued contributions.  However, the unrolled
tensor contraction code is longer which introduces some drawbacks,
such as increased C++ compile time.
\item \label{item:opt_loops}
Loop invariant code motion: A naive
implementation of a quadrature representation of
equation~\eqref{eq:weightedlaplacian,quadraturerepresentation}
in which the summations are replaced by loops results in a set
of nested loops where the number of required operations increase
exponentially with the number of loops.  However, few terms in
equation~\eqref{eq:weightedlaplacian,quadraturerepresentation} are
dependent on the summation index in each of the sums.  For instance, the
value of the function $w$ at a given quadrature point is simply computed
as $w(X^q)=\sum_{\alpha_3=1}^n \Phi(X^q)_{\alpha_3} w_{\alpha_3}$
and is therefore invariant with respect to the indices $\alpha_1$
and $\alpha_2$.  The code to compute this value can therefore be moved
outside the loop over the indices $\alpha_1$ and $\alpha_2$ This means
that the value for each entry of the element tensor $A^K_{i_{1} i_{2}}$
in equation~\eqref{eq:weightedlaplacian,quadraturerepresentation}, for
each combination of $\alpha_1$ and $\alpha_2$, can be computed in three
operations, namely a sum and two multiplications. This can be seen in
the generated code for the weighted Laplacian form which is shown in
Figure~\ref{fig:weightedlaplacian_code}.  A generic discussion of loop
invariant code motion, which is also known as `loop hoisting', can be
found in \citeN[chapter 10]{aho_ullman:dragon_book} in the context of
generic compiler optimisations.
\end{longenum}
\begin{figure}
\begin{code}{0.8}
virtual void tabulate_tensor(double* A, const double* const* w,
                             const ufc::cell& c) const
{
  ...
  // Quadrature weight
  const static double W0 = 0.5;

  // Tabulated basis functions and arrays of non-zero columns
  const static double Psi_w[1][3] =\
                   {{0.33333333333, 0.33333333333, 0.33333333333}};
  const static double Psi_vu[1][2] = {{-1, 1}};
  static const unsigned int nzc0[2] = {0, 1};
  static const unsigned int nzc1[2] = {0, 2};

  // Geometry constants
  const double G0 = Jinv_00*Jinv_10*W0*det;
  const double G1 = Jinv_01*Jinv_11*W0*det;
  const double G2 = Jinv_00*Jinv_00*W0*det;
  const double G3 = Jinv_01*Jinv_01*W0*det;
  const double G4 = Jinv_10*Jinv_10*W0*det;
  const double G5 = Jinv_11*Jinv_11*W0*det;

  // Loop integration points
  for (unsigned int ip = 0; ip < 1; ip++)
  {
    // Compute function value
    double F0 = 0;
    for (unsigned int r = 0; r < 3; r++)
      F0 += Psi_w[ip][r]*w[0][r];
    const double Gip0 = (G0 + G1)*F0;
    const double Gip1 = (G2 + G3)*F0;
    const double Gip2 = (G4 + G5)*F0;
    for (unsigned int i = 0; i < 2; i++)
    {
      for (unsigned int j = 0; j < 2; j++)
      {
        A[nzc0[i]*3 + nzc0[j]] += Psi_vu[ip][i]*Psi_vu[ip][j]*Gip1;
        A[nzc0[i]*3 + nzc1[j]] += Psi_vu[ip][i]*Psi_vu[ip][j]*Gip0;
        A[nzc1[i]*3 + nzc0[j]] += Psi_vu[ip][i]*Psi_vu[ip][j]*Gip0;
        A[nzc1[i]*3 + nzc1[j]] += Psi_vu[ip][i]*Psi_vu[ip][j]*Gip2;
      }
    }
  }
}
\end{code}
\caption{Part of the generated code for the weighted Laplacian using linear
         elements in two dimensions. The variables like \texttt{Jinv\_00} are
         components of the inverse of the Jacobian matrix and \texttt{det}
         is the determinant of the Jacobian. \texttt{A} holds the values of the
         local element tensor and \texttt{w} contains nodal values of
         the weighting function~$w$.}
\label{fig:weightedlaplacian_code}
\end{figure}

The optimisations described above take place at the final stage of the
code generation process where any given form is represented as simple
loop and algebra instructions. Therefore, the optimisations are general
and apply to all forms and elements that can be handled by \ffc{}.

The generated code for the weighted Laplacian form, using
linear Lagrange basis functions in two dimensions, is shown in
Figure~\ref{fig:weightedlaplacian_code} and demonstrates the three
optimisations described above.  The values of basis functions have been
tabulated in the variables \texttt{Psi\_w}, which is the basis for the
function $w$, and \texttt{Psi\_vu}, which contains derivatives of the
basis for the test function $v$ and the trial function~$u$.  A zero in
the table \texttt{Psi\_vu} has been eliminated, which reduces the size
of the loops over $i$ and $j$, corresponding to $i_{1}$ and $i_{2}$ in
equation~\eqref{eq:weightedlaplacian,quadraturerepresentation},
from three to two.  Note also that for each
combination of $\alpha_{1}$ and $\alpha_{2}$ in
equation~\eqref{eq:weightedlaplacian,quadraturerepresentation}, we
can evaluate the expression using only three operations.  Therefore,
increasing the number of functions and derivatives in the form will in
general not lead to an as dramatic increase in the form representation
complexity compared to the tensor contraction representation, although
additional functions might lead to an increase in the number of quadrature
points needed for exact integration.  While the above optimisations are
straightforward for simple forms and elements, their implementation
using conventional programming approaches requires manual inspection
of the form and the basis. This is often done in specialised codes,
but the extension to non-trivial forms is difficult, time consuming and
error prone. Furthermore, the optimised code may bear little relation
to the mathematical problem at hand. This makes maintenance and re-use
of the hand-generated code problematic.

Our early attempts at generating code for the quadrature representation
employed only the tabulation of basis functions as an optimisation
strategy and led to disappointing performance results.  Adding a run time
test for operations on zeros led to a performance increase, but also led
to a significant increase in the time required for the C++ compilation
of the generated code. \emph{A priori} elimination of operations on
zeroes yielded run time improvements and a significant reduction in the
time required to compile the generated code.  For complicated forms,
it was the optimisation of loops that led to dramatic performance
improvements in the run time performance. With the optimisation of
the loops, for complicated forms we have observed improvements in the
run time performance of several orders of magnitude over automatically
generated code which did not optimise the quadrature loops.

As stated in Section~\ref{subsec:quadrature}, a quadrature scheme based
on the Gauss-Legendre-Jacobi rule is applied. A further optimisation
would be to construct the quadrature rules directly for simplices, in
which case the required number of integration points for exact quadrature
could be reduced, although we recall that a user may specify the number
of integration points to be used.
%------------------------------------------------------------------------------
\section{Performance comparisons}
\label{sec:benchmarks}
We compare now generated tensor contraction and quadrature-based code
in terms of the metrics outlined in the Introduction, namely the run
time performance, the size of generated code and the speed of the
code generation phase.  The aim is to elucidate features of the two
representations for various problems with the goal of finding a guiding
principle for selecting the most appropriate representation for a given
problem.

We set the scene by first considering some typical forms of differing
complexity and nature to illustrate some trends and differences between
the representations. We then proceed with a systematic comparison
using some very simple forms for which we expect the tensor contraction
representation to prove superior, before increasing the complexity of
the forms in order to investigate the cross-over point at which the
quadrature representation becomes the better representation in terms of
run time performance.  Exact quadrature is used for all examples.

All tests were performed on an Intel Core 2 X6800 CPU at 2.93GHz with
3.2GB of RAM running Ubuntu 8.04.1 with Linux kernel 2.6.24. We used
Python version 2.5.2 and NumPy version 1.0.4 (both pertinent to \ffc{}),
and g++ version 4.2.3 with the `-O2' optimisation flag to compile
the generated C++ code which is compliant with \ufc{} version 1.1.
For tests which involve compressed sparse matrices, we use \dolfin{}
version 0.8.1 to assemble the global sparse matrix. \dolfin{} provides
various linear algebra backends, and we have used PETSc \cite{petsc:www}
as the backend for the assembly tests.  The non-zero structure of the
compressed sparse matrix is initialised and no special reordering of
degrees of freedom has been used in the assembly tests. The computer
code for the tests in this section is available \cite{oelgaard:2009}.

%------------------------------------------------------------------------------
\subsection{Performance for a selection of forms}
We set out by comparing the two representations to demonstrate the
strengths and weaknesses for different `real' forms.  The first
form considered is a mixed Poisson formulation using fifth-order
Brezzi-Douglas-Marini (BDM) elements \cite{brezzi:book}, automation
aspects of which have been addressed by \citeN{rognes:2009}.  The bilinear
form, which leads to the finite element stiffness matrix, for the mixed
Poisson problem reads
\begin{equation}
  a(\tau, w; \sigma, u) = \int_{\Omega} \tau \cdot \sigma
  - (\nabla \cdot \tau) \, u
  + w \, (\nabla \cdot \sigma) \dx,
\label{eqn:a_mixed_poisson}
\end{equation}
where $\tau, \sigma \in V$, $w, u \in W$ and
\begin{align}
  V &= \bracc{\tau \in H\brac{{\rm div}, \Omega}:
    \tau\vert_{K} \in {\rm BDM}_{k}\brac{K} \, \forall K \in \ \mathcal{T}},
  \\
  W &= \bracc{w \in L^{2}\brac{\Omega}:
    w\vert_{K} \in P_{k-1}\brac{K} \, \forall K \in \ \mathcal{T}}.
\end{align}
The \ffc{} input for this form with $k=5$ is shown in
Figure~\ref{fig:mixedpoisson_input}.
\begin{figure}
\begin{code}{0.8}
BDM = FiniteElement("Brezzi-Douglas-Marini", "triangle", 5)
DG  = FiniteElement("Discontinuous Lagrange", "triangle", 5 - 1)

mixed_element = BDM + DG

(tau, w)   = TestFunctions(mixed_element)
(sigma, u) = TrialFunctions(mixed_element)

a = (dot(tau, sigma) - div(tau)*u + w*div(sigma))*dx
\end{code}
\caption{\ffc{} input for the stiffness matrix of the mixed Poisson problem using BDM
         elements of order five~\eqref{eqn:a_mixed_poisson}.}
\label{fig:mixedpoisson_input}
\end{figure}
We also consider the generation of code for a discontinuous Galerkin
formulation of the biharmonic equation with Lagrange basis functions which
involves both cell and interior facet integrals \cite{oelgaard:2008}.
The bilinear form for this problem reads
\begin{multline}
  a(v,u) =
      \int_\Omega \nabla^{2} v  \nabla^{2} u \; \dx
    - \int_{\Gamma^{0}} \jump{\nabla v} \cdot \avg{\nabla^{2} u} \; \ds
    - \int_{\Gamma^{0}} \avg{\nabla^{2} v} \cdot \jump{\nabla u} \; \ds \\
    + \int_{\Gamma^{0}} \frac{\alpha}{h} \jump{\nabla v} \cdot \jump{\nabla u} \; \ds,
\label{eqn:a_dg_biharmonic}
\end{multline}
where the functions $v, u \in V$ and
\begin{equation}
V = \bracc{v \in H^{1}_{0}\brac{\Omega}: v_{K} \in P_{k}\brac{K} \,
  \forall K \in \mathcal{T}},
\end{equation}
and $\Gamma^{0}$ denotes all interior facets, $\alpha > 0$
is a parameter and $h$ is a measure of the cell size.  The form
compiler input for this bilinear form for the case $k=3$ is shown in
Figure~\ref{fig:biharmonicequation_input}.
\begin{figure}
\begin{code}{0.8}
element = FiniteElement("Lagrange", "triangle", 3)

v = TestFunction(element)
u = TrialFunction(element)

n = FacetNormal("triangle")
h = MeshSize("triangle")

alpha = 10.0

a = dot(div(grad(v)), div(grad(u)))*dx \
   - dot(avg(div(grad(v))), jump(grad(u), n))*dS \
   - dot(jump(grad(v), n), avg(div(grad(u))))*dS \
   + alpha/h('+')*dot(jump(grad(v),n), jump(grad(u),n))*dS
\end{code}
\caption{\ffc{} input for the stiffness matrix of a discontinuous Galerkin
         formulation for the biharmonic equation in two-dimensional
         elements of order three~\eqref{eqn:a_dg_biharmonic}.}
\label{fig:biharmonicequation_input}
\end{figure}
The third example is a complicated form which has arisen in
modelling temperature-dependent multiphase flow through porous
media~\cite{wells:2008}.  It comes from the approximate linearisation
of a stabilised finite element formulation for a particular problem
and is characterised by standard Lagrange basis functions of low order
but the products of many functions from a number of different spaces.
The physical significance of the equation is unimportant in the context
of this work, therefore it is presented in an abstract form.  The bilinear
form reads:
\begin{multline}
  a(q, p) =  \int_{\Omega} q \brac{f_0  g_2g_3/g_4} p
  -  q \brac{(1-g_{5}) \sum_{i=0}^{2} g_{i} \vect{u}_{i} \cdot \nabla p}
\\
 -  \brac{g_{6} (1-g_{5}) \sum_{i=0}^{2} \brac{f_{2i+1} / f_{2i +2}}} \nabla q \cdot \nabla p
%\\
%
 + \brac{g_{7} \sum_{i=0}^{2} g_{i} \vect{u}_{i} \cdot \nabla q} \brac{g_{3}f_{0}g_{2}/g_{4}} p
\\
 - \brac{g_{7} \sum_{i=0}^{2} g_{i} \vect{u}_{i} \cdot \nabla q}
                \brac{(1-g_{5})\sum_{i=0}^{2} g_{i} \vect{u}_{i} \cdot \nabla p}
\\
 - \brac{g_{7} \sum_{i=0}^{2} g_{i} \vect{u}_{i} \cdot \nabla q}
             \brac{ g_{6} (1-g_{5})\sum_{i=0}^{2} (f_{2i+1}/f_{2i+2}) \nabla^{2} p}
\dx,
\label{eqn:a_pressure_equation}
\end{multline}
where the test and trial functions $q, p \in V$ with
\begin{equation}
V = \bracc{v \in H^{1}\brac{\Omega}: v_{K} \in P_{2}\brac{K} \,
  \forall K \in \mathcal{T}},
\end{equation}
and the
functions $f_{i} \in V_{f}$, $g_{i}\in V_{g}$ and $\vect{u}_{i}\in V_{u}$
are coefficient functions.
The coefficients spaces are:
\begin{align}
 V_{f} &= \bracc{f \in H^{1}\brac{\Omega}: f_{K} \in P_{1}\brac{K} \,
  \forall K \in \mathcal{T}},
\\
  V_{g} &= \bracc{g \in L^{2}\brac{\Omega}: g_{K} \in P_{1}\brac{K} \,
  \forall K \in \mathcal{T}},
\\
  V_{u} &= \bracc{\vect{u} \in \brac{L^{2}\brac{\Omega}}^{2}: \vect{u}_{K} \in \brac{P_{1}\brac{K}}^{2} \,
  \forall K \in \mathcal{T}}.
\end{align}
The coefficient functions are either prescribed or come from the solution
of other equations.  The input to the compiler for this form is shown
in Figure~\ref{fig:pressureequation_input}.  Due to the origins of this
form, we denote it informally as the `pressure equation'.
\begin{figure}
\begin{code}{0.8}
scalar_p = FiniteElement("Lagrange","triangle",2)
scalar   = FiniteElement("Lagrange","triangle",1)
dscalar  = FiniteElement("Discontinuous Lagrange","triangle",0)
vector   = VectorElement("Discontinuous Lagrange", "triangle", 1)

q   = TestFunction(scalar_p)
p   = TrialFunction(scalar_p)

f0, f1, f2, f3, f4, f5, f6 = [Function(scalar) for i in range(7)]
g0, g1, g2, g3, g4, g5, g6, g7 = [Function(dscalar) for i in range(8)]
u0, u1, u2 = [Function(vector) for i in range(3)]

Sgu = mult(g0, u0) + mult(g1, u1) + mult(g2, u2)
S   = g6*(1 - g5)*(f1/f2 + f3/f4 + f5/f6)

a_0 = q*g3*f0*g2/g4*p\
    - q*(1 - g5)*dot(Sgu, grad(p))\
    - S*dot(grad(q), grad(p))

a_1 = g7*dot(Sgu, grad(q))*g3*f0*g2/g4*p\
    - g7*dot(Sgu, grad(q))*(1 - g5)*dot(Sgu, grad(p))\
    + g7*dot(Sgu, grad(q))*S*div(grad(p))

a = (a_0 + a_1)*dx
\end{code}
\caption{\ffc{} input for the `pressure equation' in two
dimensions~\eqref{eqn:a_pressure_equation}.}
\label{fig:pressureequation_input}
\end{figure}

The three forms have been compiled with \ffc{} using
the tensor contraction and quadrature representations. In
Table~\ref{tab:various_forms_generation}, the time required to generate
the code, the size of the generated code and the time required to compile
the C++ code are reported for each form. Results are presented for the
tensor contraction case, together with the ratio of the time/size for the
quadrature representation case divided by the time/size required for the
tensor contraction representation case, denoted by~q/t.  In measuring
the C++ compile time and the run time performance, the generated code
has been compiled against the library \dolfin{}.
\begin{table}
\caption{Timings and code size for the compilation phase for the various
variational forms.
  `generation' is the time required by \ffc{} to generate the tensor
  contraction code; `size' is the size of the generated tensor contraction
  code; and `C++' is the time to compile the generated C++ code. The
     ratio q/t is the ratio between quadrature and tensor contraction
     representations.}
\label{tab:various_forms_generation}
\begin{center}
\begin{tabular}{l|rr|rr|rr}
Form              & generation {\scriptsize [s]} & q/t   & size {\scriptsize [kB]} & q/t  & C++  {\scriptsize [s]}  & q/t  \\[0.5ex]
\hline
mixed Poisson     &  3.9                   & 1.00  & 1500                    & 0.92 & 15.7                    & 0.76 \\[1ex]
DG biharmonic     & 16.2                   & 0.35  & 3200                    & 0.13 & 47.5                    & 0.19 \\[1ex]
pressure equation & 35.1                   & 1.03  & 2600                    & 0.19 & 41.4                    & 0.22
\end{tabular}
\end{center}
\end{table}
Noteworthy from the results in Table~\ref{tab:various_forms_generation}
is that the generation phase for the quadrature representation is
at least as fast as the tensor contraction representation generation
phase (the difference for the pressure equation is only slight). Our
experience is that the difference grows in favour of the quadrature
representation for complicated forms.  In all cases the size of the
generated quadrature code is smaller than the tensor contraction code,
which is reflected in the C++ compile time. The differences in the C++
compile time are substantial for the biharmonic and pressure equations
(approximately a factor of five), which is important during the code
development phase with frequent recompilations.

Timings and operation counts for the three forms are presented in
Table~\ref{tab:various_forms_flops}.
\begin{table}
\caption{Run time performance for the various variational forms.}
\label{tab:various_forms_flops}
\begin{center}
\begin{tabular}{l|rr|rr}
Form              &  flops   & q/t   & run time  {\scriptsize [s]} & q/t \\[0.5ex]
\hline
mixed Poisson     &   12866  & 73.90 &   4.5                      & 60.33\\[1ex]
DG biharmonic     &   26420  &  1.19 &  25.0                      &  1.27\\[1ex]
pressure equation &  160752  &  0.17 & 190.0                      &  0.17\\
\end{tabular}
\end{center}
\end{table}
We define the number of floating point operations (flops) as the sum of
all `+' and `$\ast$' operators in the code for computing the element
matrix.  Although multiplications are generally more expensive than
additions, this definition provides a good measure for the performance
of the generated code.  The compound operator `+=' is counted as one
operation.  For the run time performance, the time required to compute the
element tensors $N$ times is recorded. For the mixed Poisson problem $N =
4.5 \times 10^{5}$, for the discontinuous Galerkin biharmonic problem
$N = 1 \times 10^{6}$ and for the pressure equation $N = 2.5 \times
10^{6}$.  Table~\ref{tab:various_forms_flops} presents the timings and
operation counts for tensor contraction representation, together with the
ratio of the quadrature representation case and the tensor contraction
representation case,~q/t.  The run time performance is indicative of an
aspect of the two representations; there can be significant performance
difference depending on the nature of the differential equation. For the
mixed Poisson problem, the tensor contraction representation is close to a
factor of sixty faster than the quadrature representation, whereas for the
pressure equation the quadrature representation is close to a factor of
six faster than the tensor contraction case.  This observation of dramatic
differences in run time performance suggests the possibility of devising a
strategy for determining the best representation, without generating the
code for each case.  Such concepts have been successfully developed in
digital signal processing \cite{pueschel:05}. For forms with a relatively
simple structure, devising such a scheme is straightforward. However,
it turns out to be non-trivial for arbitrary forms.
%------------------------------------------------------------------------------
\subsection{Performance for common, simple forms}
We consider now the performance of the two representations for two
canonical examples: the scalar `mass' matrix and the `elasticity-like'
stiffness matrix.  The input for the mass matrix form is shown in
Figure~\ref{fig:mass_input} and the input for the elasticity-like
stiffness matrix is shown in Figure~\ref{fig:elasticity_input}.
\begin{figure}
\begin{code}{0.8}
element = FiniteElement("Lagrange", "triangle", 2)

v = TestFunction(element)
u = TrialFunction(element)

a = dot(v, u)*dx
\end{code}
\caption{\ffc{} input for the mass matrix in two dimension with element order
         $q=2$.}
\label{fig:mass_input}
\end{figure}
\begin{figure}
\begin{code}{0.8}
element = VectorElement("Lagrange", "tetrahedron", 3)

v = TestFunction(element)
u = TrialFunction(element)

def eps(v):
    return grad(v) + transp(grad(v))

a = 0.25*dot(eps(v), eps(u))*dx
\end{code}
\caption{\ffc{} input for the elasticity-like matrix in three dimensions with element order $q=3$.}
\label{fig:elasticity_input}
\end{figure}
The performance of the two representations are compared for two- and
three-dimensional cases on simplicies and for various polynomial orders.
Code is generated using \ffc{}, and we report the number of floating
point operations required to form the element matrix for all cases.
In addition to reporting the number of floating point operations,
the time required to compute the element matrix $N$ times is also
presented, which we expect in most cases to be strongly correlated to
the floating point operations count.  As before, values are reported for
the tensor contraction representation case together with the ratio of the
quadrature value over the tensor contraction value.  We also report the
time required for insertion into a sparse matrix, which is independent
of the element matrix representation.  The total assembly time is the
`run time' plus the `insertion' time, which provides a picture of the
overall assembly performance.  The ratio of the total assembly time for
the quadrature representation over the total assembly time for the tensor
contraction representation, denoted by~a$_\text{q}$/a$_\text{t}$, is also
presented.  When taking this into account, for some forms the difference
in performance between different representations appears less drastic.

The various timings for the mass matrix problem are reported in
Table~\ref{tab:mass_comparison_2D} for the two-dimensional case and
in Table~\ref{tab:mass_comparison_3D} for the three-dimensional case.
What is clear from these results is that tremendous speed-ups for
computing the element matrices can be achieved using the tensor
contraction representation, particularly as the element order is
increased.  This is perhaps not surprising considering that the geometry
tensor for this case is simply a scalar, therefore the entire matrix is
essentially precomputed. The speed-up is mitigated, however, by the time
required to insert terms into a sparse matrix. For the case of $q=4$ in
three dimensions, the tensor contraction representation is a factor of 358
faster for computing the element matrix, but when insertion is included an
overall speed-up factor of only 2.76 is observed. A factor of 2.76 is not
trivial, but obviously to reap the full benefits of the tensor contraction
approach for these types of problems, matrix insertion must be addressed.
If in addition the time required to perform the remaining parts of the
finite element procedure such as mesh initialisation, application of
boundary conditions, and solving the resulting system of equations is
taken into account the q/t ratio will become even closer to unity.

The various timings for the elasticity-like stiffness matrix are presented
in Table~\ref{tab:elasticity_comparison_2D} for the two-dimensional case
and in Table~\ref{tab:elasticity_comparison_3D} for the three-dimensional
case.  Compared to the mass matrix, the differences in performance of the
tensor contraction representation relative to quadrature representation
are less dramatic, but nonetheless substantial, especially for
higher-order functions in three dimensions.
\begin{table}
\caption{Timings for the mass matrix in two dimensions for varying polynomial
         order basis~$q$.}
\label{tab:mass_comparison_2D}
\begin{center}
\begin{tabular}{l|rr|rr|r|r}
                                 & flops & q/t & run time [s] & q/t & insertion [s] & a$_\text{q}$/a$_\text{t}$\\
\hline
$q = 1$ ($N = 1 \times 10^{7}$)  &   10  &  11 & 0.13         &   6 &  8            & 1.07\\
$q = 2$ ($N = 1 \times 10^{7}$)  &   25  &  39 & 0.27         &  19 & 23            & 1.21\\
$q = 3$ ($N = 1 \times 10^{7}$)  &   89  &  54 & 0.67         &  37 & 59            & 1.40\\
$q = 4$ ($N = 1 \times 10^{6}$)  &  214  &  79 & 0.12         &  71 & 13            & 1.64\\
$q = 5$ ($N = 1 \times 10^{6}$)  &  442  & 108 & 0.20         & 112 & 24            & 1.91
\end{tabular}
\end{center}
\end{table}
\begin{table}
\caption{Timings for the mass matrix in three dimensions for varying polynomial
         order basis~$q$.}
\label{tab:mass_comparison_3D}
\begin{center}
\begin{tabular}{l|rr|rr|r|r}
                                 & flops & q/t & run time [s] & q/t & insertion [s] & a$_\text{q}$/a$_\text{t}$\\
\hline
$q = 1$ ($N = 1 \times 10^{7}$)  &   17  &  23 & 0.31         &   8 &  11           & 1.19\\
$q = 2$ ($N = 1 \times 10^{7}$)  &  101  &  80 & 0.65         &  60 &  72           & 1.52\\
$q = 3$ ($N = 1 \times 10^{6}$)  &  281  & 273 & 0.18         & 202 &  34           & 2.06\\
$q = 4$ ($N = 1 \times 10^{6}$)  & 1226  & 375 & 0.60         & 358 & 121           & 2.76
\end{tabular}
\end{center}
\end{table}
\begin{table}
\caption{Timings for the elasticity-like matrix in two dimensions for varying
         polynomial order basis~$q$.}
\label{tab:elasticity_comparison_2D}
\begin{center}
\begin{tabular}{l|rr|rr|r|r}
                                 & flops  & q/t & run time [s] & q/t & insertion [s] & a$_\text{q}$/a$_\text{t}$\\
\hline
$q = 1$ ($N = 1 \times 10^{7}$)  &    236 & 1.1 & 0.64         &  3  &  19           & 1.06\\
$q = 2$ ($N = 1 \times 10^{7}$)  &    728 &   7 & 1.65         & 17  & 100           & 1.25\\
$q = 3$ ($N = 1 \times 10^{6}$)  &   2728 &  13 & 0.69         & 33  &  29           & 1.74\\
$q = 4$ ($N = 1 \times 10^{5}$)  &   7724 &  20 & 0.41         & 22  &   7           & 2.16
\end{tabular}
\end{center}
\end{table}
\begin{table}
\caption{Timings for the elasticity-like matrix in three dimensions for varying
         polynomial order basis~$q$.}
\label{tab:elasticity_comparison_3D}
\begin{center}
\begin{tabular}{l|rr|rr|r|r}
                                 & flops  & q/t & run time [s] & q/t & insertion [s] & a$_\text{q}$/a$_\text{t}$\\
\hline
$q = 1$ ($N = 1 \times 10^{6}$)  &   1098 & 1.1 & 0.27         &   3 &  8            & 1.07\\
$q = 2$ ($N = 1 \times 10^{5}$)  &   8622 &  11 & 0.36         &  15 &  9            & 1.54\\
$q = 3$ ($N = 1 \times 10^{4}$)  &  44010 &  38 & 0.19         &  60 &  6            & 2.81\\
$q = 4$ ($N = 1 \times 10^{3}$)  & 155529 &  90 & 0.08         & 115 &  2            & 5.38
\end{tabular}
\end{center}
\end{table}
%------------------------------------------------------------------------------
\subsection{Performance for forms of increasing complexity}
The complexity of the forms investigated in the previous section is now
increased systematically in order to examine under which circumstances
the quadrature representation will be more favourable in terms of
run time performance. The comparison is based on the floating point
operation count as this is a good indicator of performance and the
size of the generated file for a large class of problems.  We consider
the `complexity' of a variational form to increase when the number of
function products increases and when the number of derivatives present
increases.  Increasing the number of derivatives and/or the numbers
of functions appearing in a form leads to higher rank tensors for the
tensor contraction representation. Also, increases in the polynomial
order of the basis of a function leads to an increase in complexity of
the geometry tensor while increases in the polynomial order of the basis
functions lead to an increase in complexity of the reference tensor. We
initially restrict ourselves to manipulating the number of function
multiplications in the forms and the polynomial order of these functions,
before introducing products of derivatives.

To generate forms of greater complexity than those in the
previous section, we take the mass matrix and elasticity-like
problems with a Lagrange basis of order $q$, and premultiply the
forms with $n_{f}$ functions of order~$p$.  An example is shown in
Figure~\ref{fig:mass_complex_input} for the mass matrix pre-multiplied
by coefficient functions where $q=2$, $n_{f}=2$ and~$p=3$.
\begin{figure}
\begin{code}{0.8}
element   = FiniteElement("Lagrange", "triangle", 2)
element_f = FiniteElement("Lagrange", "triangle", 3)

v = TestFunction(element)
u = TrialFunction(element)

f = Function(element_f)
g = Function(element_f)

a = f*g*dot(v, u)*dx
\end{code}
\caption{\ffc{} input for the mass matrix in two dimension with with $q=2$,
          premultiplied by two functions ($n_{f}=2$) of order $p=3$.}
\label{fig:mass_complex_input}
\end{figure}
A comparison of the representations for the mass matrix with a different
number of premultiplying functions and a range of orders $p$ and $q$
are presented in Table~\ref{tab:mass2D_complex_comparison} for the
two-dimensional case and in Table~\ref{tab:mass3D_complex_comparison}
for the three-dimensional case.
\begin{table}
\caption{The number of operations and the ratio between number of operations
         for the two representations for the mass matrix in two dimensions as
         a function of different polynomial orders and numbers of functions.}
\label{tab:mass2D_complex_comparison}
\begin{center}
\begin{tabular}{l|rr|rr|rr|rr}
\multicolumn{1}{c}{} & \multicolumn{2}{c}{$n_f = 1$} & \multicolumn{2}{c}{$n_f = 2$} & \multicolumn{2}{c}{$n_f = 3$} & \multicolumn{2}{c}{$n_f = 4$}\\
                  & flops & q/t                   & flops & q/t                   & flops & q/t          & flops & q/t\\
\hline
$p = 0$, $q = 1$  &   10    & 11.30               &    11   &  10.36              &     12  &  9.58               &      13 &  8.92 \\
$p = 0$, $q = 2$  &   25    & 39.28               &    26   &  37.81              &     27  & 36.44               &      28 & 35.18 \\
$p = 0$, $q = 3$  &   89    & 54.12               &    90   &  53.55              &     91  & 52.96               &      92 & 52.39 \\
$p = 0$, $q = 4$  &  214    & 78.98               &   215   &  78.61              &    216  & 78.25               &     217 & 77.90 \\
\hline %
$p = 1$, $q = 1$  &   48    &  2.91               &   171   &   2.21              &    558  &  0.79               &    1773 &  0.51 \\
$p = 1$, $q = 2$  &  183    &  5.70               &   474   &   4.15              &   1917  &  1.09               &    5448 &  0.63 \\
$p = 1$, $q = 3$  &  431    & 11.43               &  1442   &   5.46              &   5381  &  1.50               &   15368 &  0.77 \\
$p = 1$, $q = 4$  & 1128    & 15.14               &  3819   &   6.50              &  12006  &  2.09               &   36549 &  0.94 \\
\hline
$p = 2$, $q = 1$  &   81    &  4.56               &   555   &   1.56              &   4143  &  0.40               &   26007 &  0.11 \\
$p = 2$, $q = 2$  &  258    &  7.57               &  2412   &   1.40              &  14724  &  0.36               &   94428 &  0.08 \\
$p = 2$, $q = 3$  &  950    &  8.26               &  6800   &   1.73              &  42998  &  0.39               &  251876 &  0.10 \\
$p = 2$, $q = 4$  & 2457    & 10.10               & 15987   &   2.15              &  95247  &  0.48               &  585567 &  0.10 \\
\hline
$p = 3$, $q = 1$  &  181    &  2.44               &  1715   &   1.02              &  20991  &  0.16               &  218767 &  0.03 \\
$p = 3$, $q = 2$  &  550    &  3.78               &  6992   &   0.78              &  73596  &  0.11               &  754084 &  0.02 \\
$p = 3$, $q = 3$  & 1910    &  4.21               & 20100   &   0.84              & 202900  &  0.11               & 2038820 &  0.02 \\
$p = 3$, $q = 4$  & 4285    &  5.86               & 44099   &   1.04              & 452775  &  0.13               & 4538983 &  0.02
\end{tabular}
\end{center}
\end{table}
\begin{table}
\caption{The number of operations and the ratio between number of operations
         for the two representations for the mass matrix in three dimensions as
         a function of different polynomial orders and numbers of functions.}
\label{tab:mass3D_complex_comparison}
\begin{center}
\begin{tabular}{l|rr|rr|rr|rr}
\multicolumn{1}{c}{} & \multicolumn{2}{c}{$n_f = 1$} & \multicolumn{2}{c}{$n_f = 2$} & \multicolumn{2}{c}{$n_f = 3$} & \multicolumn{2}{c}{$n_f = 4$}\\
                  & flops & q/t          & flops & q/t          & flops & q/t          & flops & q/t\\
\hline
$p = 1$, $q = 1$  &   116   &   4.00              &    528  &   3.43              &    2224 &  0.92               &    9200 &  0.59 \\
$p = 1$, $q = 2$  &   608   &  13.77              &   3084  &   6.62              &   12412 &  1.69               &   52124 &  0.81 \\
$p = 1$, $q = 3$  &  2660   &  29.11              &  12432  &  12.26              &   46528 &  3.30               &  205424 &  1.30 \\
$p = 1$, $q = 4$  &  7955   &  57.90              &  38007  &  20.99              &  155751 &  5.14               &  622679 &  2.04 \\
\hline
$p = 2$, $q = 1$  &   314   &   6.02              &   3336  &   1.75              &   34984 &  0.40               &  359984 &  0.08 \\
$p = 2$, $q = 2$  &  1838   &  11.21              &  20100  &   2.13              &  202900 &  0.39               & 2034140 &  0.06 \\
$p = 2$, $q = 3$  &  7610   &  20.07              &  79800  &   3.36              &  765592 &  0.57               & 8039600 &  0.08 \\
$p = 2$, $q = 4$  & 23285   &  34.29              & 239415  &   5.33              & 2451775 &  0.78               &24538775 &  0.11 \\
\hline
$p = 3$, $q = 1$ &   644   &   3.77              &  13584  &   1.21              &  279984 &  0.13                & 5759984 &  0.02 \\
$p = 3$, $q = 2$ &  3752   &   5.83              &  80700  &   1.03              & 1564572 &  0.09                &  \multicolumn{2}{c}{\tt \ffc{} failure} \\
$p = 3$, $q = 3$ & 14684   &  10.57              & 315216  &   1.40              & 6372120 &  0.11                &  \multicolumn{1}{c}{-} & \multicolumn{1}{c}{-} \\
$p = 3$, $q = 4$ & 47795   &  16.80              & 979575  &   1.96              &19594199 &  0.14                &  \multicolumn{1}{c}{-} & \multicolumn{1}{c}{-}
\end{tabular}
\end{center}
\end{table}
What is clear from Table~\ref{tab:mass2D_complex_comparison}
is that with few premultiplying functions, the tensor contraction
approach is generally more efficient, even for relatively high order
premultiplying functions.  The situation changes quite dramatically
for $p > 0$ as the number of premultiplying functions increases, and
as the polynomial order of the premultiplying functions increases.
The cases with numerous premultiplying functions are typical of the
Jacobian resulting from the linearisation of a nonlinear differential
equation in a practical simulation, and are therefore important.  It is
also noted that the tensor contraction representation is more efficient
for increases in $q$, however, this effect is less pronounced for the
cases where $n_f > 1$ and $p > 0$.  Obviously, the selection of the
representation can have a tremendous performance impact.  The relative
performance of the representations in three dimensions is shown in
Table~\ref{tab:mass3D_complex_comparison}. The number of operations
has increased relative to the two-dimensional case, which corresponds
to an increase in the size of the generated code.  For the more complex
forms, compilation of the generated C++ code for the tensor contraction
representation is no longer feasible, and in some cases simply not
possible due to compiler limitations.  For the most complicated cases,
\ffc{} was unable to generate tensor contraction code due to memory
being exhausted. In practice, time is the limiting factor as a memory
error is usually only encountered after many hours of code generation
for the tensor contraction case.  \ffc{} was able to generate quadrature
representation code for all cases.

Interestingly, for complicated forms the operation count is not always
a good indicator of performance. For the three-dimensional mass matrix
case with $p=1$, $q=4$ and $n_{f} = 4$, we would expect from the operation
count that the tensor contraction representation would be faster. However,
when computing the element tensor $48000$ times, we observed a ratio of
$\text{q}/\text{t} = 0.78$, indicating that the quadrature representation
is faster.  Noteworthy for this case is that the size of the generated
code for tensor contraction representation is 11~MB, while the size of
the generated quadrature code is only~362~kB. This size difference leads
not only to a significant difference in the C++ compile time, but also
appears to result in a drop in run time performance.  The performance
drop could be attributed to the increased memory traffic noted by
\citeN{kirby:2006c}. Also, it may be that the compiler is unable to
perform effective optimisations on the unrolled code, or that the
compiler is particularly effective at optimising the loops in the
generated quadrature code.

A similar comparison is made for
elasticity-like forms and the results are presented in
Table~\ref{tab:elasticity2D_complex_comparison} for the two-dimensional
case and in Table~\ref{tab:elasticity3D_complex_comparison} for the
three-dimensional case.
\begin{table}
\caption{The number of operations and the ratio between number of operations
         for the two representations for the elasticity-like tensor in two
         dimensions as a function of different polynomial orders and numbers of
         functions.}
\label{tab:elasticity2D_complex_comparison}
\begin{center}
\begin{tabular}{l|rr|rr|rr}
\multicolumn{1}{c}{} & \multicolumn{2}{c}{$n_f = 1$} & \multicolumn{2}{c}{$n_f = 2$} & \multicolumn{2}{c}{$n_f = 3$}\\
                  & flops & q/t          & flops & q/t          & flops & q/t\\
\hline
$p = 1$, $q = 1$  &    888  &  0.34               &    3060 &  0.36               &   10224 & 0.11\\
$p = 1$, $q = 2$  &   3564  &  1.42               &   11400 &  1.01               &   35748 & 0.33\\
$p = 1$, $q = 3$  &  10988  &  3.23               &   34904 &  1.82               &  100388 & 0.63\\
$p = 1$, $q = 4$  &  26232  &  5.77               &   82548 &  2.87               &  254304 & 0.93\\
\hline
$p = 2$, $q = 1$  &    888  &  1.20               &    8220 &  0.31               &   54684 & 0.09\\
$p = 2$, $q = 2$  &   7176  &  1.59               &   41712 &  0.49               &  284232 & 0.11\\
$p = 2$, $q = 3$  &  22568  &  2.80               &  139472 &  0.71               &  856736 & 0.17\\
$p = 2$, $q = 4$  &  54300  &  4.36               &  337692 &  1.01               & 2058876 & 0.23\\
\hline
$p = 3$, $q = 1$  &   3044  &  0.36               &   30236 &  0.16               &  379964 & 0.02\\
$p = 3$, $q = 2$  &  12488  &  0.92               &  126368 &  0.26               & 1370576 & 0.03\\
$p = 3$, $q = 3$  &  36664  &  1.73               &  391552 &  0.37               & 4034704 & 0.05\\
$p = 3$, $q = 4$  &  92828  &  2.55               &  950012 &  0.49               & 9566012 & 0.06\\
\hline
$p = 4$, $q = 1$  &   3660  &  0.68               &   73236 &  0.11               & 1275624 & 0.01\\
$p = 4$, $q = 2$  &  17652  &  1.16               &  296712 &  0.16               & 4628460 & 0.02\\
$p = 4$, $q = 3$  &  57860  &  1.71               &  903752 &  0.22               &13716836 & 0.02\\
$p = 4$, $q = 4$  & 138984  &  2.46               & 2133972 &  0.29               &32289984 & 0.03
\end{tabular}
\end{center}
\end{table}
\begin{table}
\caption{The number of operations and the ratio between number of operations
         for the two representations for the elasticity-like tensor in three
         dimensions as a function of different polynomial orders and numbers of
         functions.}
\label{tab:elasticity3D_complex_comparison}
\begin{center}
\begin{tabular}{l|rr|rr|rr}
\multicolumn{1}{c}{}                  & \multicolumn{2}{c}{$n_f = 1$} & \multicolumn{2}{c}{$n_f = 2$} & \multicolumn{2}{c}{$n_f = 3$}\\
                  & flops & q/t          & flops & q/t          & flops & q/t\\
\hline
$p = 1$, $q = 1$  &    5508 &  0.26               &   25200 &  0.40               &  112176 & 0.09\\
$p = 1$, $q = 2$  &   40176 &  2.42               &  169020 &  1.95               &  597564 & 0.55\\
$p = 1$, $q = 3$  &  201348 &  8.37               &  735408 &  5.44               & 3422160 & 1.16\\
$p = 1$, $q = 4$  &  708291 & 19.78               & 2958831 &  9.25               &11728143 & 2.33\\
\hline
$p = 2$, $q = 1$  &   13986 &  0.70               &  158256 &  0.22               & 1691676 & 0.05\\
$p = 2$, $q = 2$  &  103518 &  3.17               & 1059804 &  0.74               &11132244 & 0.14\\
$p = 2$, $q = 3$  &  450882 &  8.86               & 5417136 &  1.44               & \multicolumn{2}{c}{\tt \ffc{} failure} \\
$p = 2$, $q = 4$  & 1836225 & 14.90               &18941967 &  2.50               & \multicolumn{1}{c}{-} & \multicolumn{1}{c}{-} \\
\hline
$p = 3$, $q = 1$  &   11160 &  0.89               &  443376 &  0.19               &13218516 & 0.01\\
$p = 3$, $q = 2$  &  186624 &  1.76               & 4402620 &  0.35               & \multicolumn{2}{c}{\tt \ffc{} failure} \\
$p = 3$, $q = 3$  & 1035684 &  3.86               &21777552 &  0.62               & \multicolumn{1}{c}{-} & \multicolumn{1}{c}{-} \\
$p = 3$, $q = 4$  & 3681171 &  7.43               & \multicolumn{2}{|c|}{\tt \ffc{} failure} & \multicolumn{1}{c}{-} & \multicolumn{1}{c}{-} \\
\hline
$p = 4$, $q = 1$  &   49311 &  0.69               & 1940256 &  0.09               & \multicolumn{2}{c}{\tt \ffc{} failure} \\
$p = 4$, $q = 2$  &  364275 &  2.14               &13527684 &  0.20               & \multicolumn{1}{c}{-} & \multicolumn{1}{c}{-}
\end{tabular}
\end{center}
\end{table}
Similar trends to those observed for the mass matrix hold. In three
dimensions \ffc{} fails to generate code for a number of the more complex
forms using the tensor contraction representation. Code generation
using the quadrature representation is successful in all cases.  Also,
file size considerations, especially in the three-dimensional cases,
will rule out the tensor contraction representation for a number of
forms where, based on the ratio, it would be expected to outperform the
quadrature representation.  It is more difficult in these cases to make
broad generalisation as to the best representation. This again suggests
that a method for automatically determining the best representation
based on inspection of the form may be interesting.

Finally, we investigate the influence of premultiplying a
vector-valued Poisson form by the divergence of vector-valued
functions. The form for the case $n_{f}=2$, $p=3$ and $q=2$ is shown
in Figure~\ref{fig:vector_poisson_complex_input}.
\begin{figure}
\begin{code}{0.8}
element   = VectorElement("Lagrange", "triangle", 2)
element_f = VectorElement("Lagrange", "triangle", 3)

v = TestFunction(element)
u = TrialFunction(element)

f = Function(element_f)
g = Function(element_f)

a = div(f)*div(g)*dot(grad(v), grad(u))*dx
\end{code}
\caption{\ffc{} input for the vector-valued Poisson problem in two dimension
          with with $q=2$, premultiplied by the divergence of two vector valued
          functions ($n_{f}=2$) of order $p=3$.}
\label{fig:vector_poisson_complex_input}
\end{figure}
A comparison of tensor contraction and quadrature representations
is performed, as in the previous cases, and the results are shown in
Table~\ref{tab:function_derivatives2D_comparison}.
\begin{table}
\caption{The number of operations and the ratio between number of operations
         for the two representations for the vector-valued Poisson problem in
         two dimensions as a function of different polynomial orders and numbers
         of functions.}
\label{tab:function_derivatives2D_comparison}
\begin{center}
\begin{tabular}{l|rr|rr}
\multicolumn{1}{c}{} & \multicolumn{2}{c}{$n_f = 1$} & \multicolumn{2}{c}{$n_f = 2$} \\
                  & flops  & q/t                  & flops & q/t    \\
\hline
$p = 1$, $q = 1$  &    686 & 0.33                 &    6126 & 0.07 \\
$p = 1$, $q = 2$  &   2180 & 1.22                 &   18372 & 0.18 \\
$p = 1$, $q = 3$  &   8068 & 2.23                 &   66372 & 0.29 \\
$p = 1$, $q = 4$  &  22526 & 3.38                 &  183870 & 0.43 \\
\hline
$p = 2$, $q = 1$  &   1390 & 0.17                 &   24558 & 0.06 \\
$p = 2$, $q = 2$  &   7768 & 0.36                 &  162744 & 0.05 \\
$p = 2$, $q = 3$  &  24872 & 0.73                 &  512872 & 0.07 \\
$p = 2$, $q = 4$  &  60190 & 1.27                 & 1246478 & 0.10 \\
\hline
$p = 3$, $q = 1$  &   2094 & 0.42                 &   96750 & 0.04 \\
$p = 3$, $q = 2$  &  11640 & 0.55                 &  541800 & 0.03 \\
$p = 3$, $q = 3$  &  44776 & 0.73                 & 1697576 & 0.03 \\
$p = 3$, $q = 4$  & 110774 & 1.09                 & 4099406 & 0.04
\end{tabular}
\end{center}
\end{table}
Premultiplying forms with derivatives of functions clearly increases the
complexity to such a degree that the tensor contraction representation
involves fewer operations for only a very limited number of the considered
cases.
%------------------------------------------------------------------------------
\section{Conclusions}
\label{sec:conclusions}
We have presented two representations, namely the tensor contraction
and quadrature representations, for the computation of element tensors
arising in the finite element method. The generation of code for these
representations is automated and permits both the rapid development of
solvers for broad classes of problems and the application of specialised
performance optimisations. In particular, we have presented strategies
for optimising automatically the quadrature representation code which are
manifest in the form compiler FFC.  The strategies introduce negligible
overhead in the code generation phase, but can yield substantial run
time speed-ups over non-optimised code.  The presented techniques are
possible with conventional `hand' coding, and in fact commonly employed
in specialised codes and simple problems.  Automation makes the approach
generic and allows the application of simple but tedious to implement
by hand strategies to an unlimited range of problems. Our efforts
to develop automated optimisations are based on the experience that
non-optimised generated code is not competitive in terms of performance.
For the problems which we have investigated, the automated optimisations
employed in the generated code are sufficient that insertion of entries
into a space matrix is often the limiting factor in the assembly of
sparse global finite element matrices.

The relative performance of two representations of finite element tensors
has been investigated for a range of different problems. In assessing
performance we have considered the time to generate the code, the number
of floating point operations involved in the generated code, the time
to compile the low-level generated code and the runtime for assembly.
Numerical experiments have shown that the relative performance of the
two representations in terms of the aforementioned measures can differ
substantially depending on the nature of the considered variational form.
We have observed that for some forms the tensor contraction involves
of the order 300 times fewer operations relative to the quadrature
representation to construct the element matrix, whereas for other examined
forms the quadrature representation requires of the order 100 times
fewer operations. In general, the tensor contraction approach deals well
with forms which involve high-order bases and few coefficient functions,
whereas the quadrature representation is more efficient as the number of
coefficient functions (other than constants coefficients) and derivatives
in a form increases.  Hence, in general the quadrature representation is
significantly faster for more complicated forms. Also, relative to the
tensor contraction representation, the time required to generate code for
the quadrature representation is less and the size of the generated code
is smaller. For forms which involve a number of coefficient functions,
the code generation phase is faster for the quadrature representation.
For some of the examined examples, the size of the generated code file
is close to a factor of 10 smaller for the quadrature representation
relative to the tensor contraction representation.  This impacts heavily
on the time required to compile the generated C++ code.

Automation is most attractive for complicated forms as they are time
consuming to implement, implementations are error prone and performance
is more elusive. Therefore, in addressing the quadrature representation
in the context of automated code generation, we have extended the
applicability of automated modelling and of \ffc{} to more complicated
variational forms.  In practice, a sophisticated solver will often
involve the assembly of various forms of differing complexity, so having
both tensor contraction and quadrature representations as part of the
computational arsenal allows the most appropriate representation for a
given form to be used.  An interesting point is the automated selection
of the best representation.  \ffc{} presently computes the operation
count for the code which is generated, on the basis of which a choice
could be made, but this involves generating computer code for each case
which can be time consuming.  Ideally, the form compiler would select
the best representation based on an inspection of the form. It turns
out, however, that this is a non-trivial task if the goal is a general
approach which holds for any form which \ffc{} can handle and is a topic
of ongoing investigation.
%------------------------------------------------------------------------------
\begin{ack}
KB{\O} acknowledges the support of the Netherlands Technology Foundation
STW, the Netherlands Organisation for Scientific Research and the Ministry
of Public Works and Water Management. Valuable discussions with Anders
Logg are gratefully acknowledged.
\end{ack}
%------------------------------------------------------------------------------
\bibliographystyle{acmtrans}
\bibliography{references}
%------------------------------------------------------------------------------
\begin{received}
\end{received}
%------------------------------------------------------------------------------
\end{document}